\documentclass[11pt]{article}
\usepackage{amssymb}
\usepackage{amsmath}
\usepackage{amscd}
\usepackage{latexsym}

\catcode`@=11
\renewcommand{\section}{\@startsection{section}{1}{0pt}{\medskipamount}
{\medskipamount}{\large\bf}}
\numberwithin{equation}{section}
\catcode`@=12
\def\a{\alpha}
\def\b{\beta}
\def\g{\gamma}

\def\de{\delta}
\def\ve{\varepsilon}
\def\h{\eta}
\def\th{\theta}

\def\la{\lambda}
\def\m{\mu}

\def\vp{\varphi}

\def\j{\psi}

\def\da{\dot\alpha}
\def\db{\dot\beta}
\def\1{\dot 1}
\def\2{\dot 2}
\def\Ups{\Upsilon}
\def\eps{\epsilon}

\newcommand{\C}{\mathbb C}
\newcommand{\R}{\mathbb R}

\newcommand{\Acal}{{\cal A}}
\newcommand{\Ncal}{{\cal N}}
\newcommand{\Lcal}{{\cal L}}

\def\N2{$N{=}2$}
\def\pa{\mbox{$\partial$}}
\def\diff{\mbox{d}}
\def\tr{{\rm tr}}
\def\sfrac#1#2{{\textstyle\frac{#1}{#2}}}
\def\>{\rangle}
\def\<{\langle}
\def\+{\dagger}
\def\={\ =\ }

\begin{document}

\begin{titlepage}
\setcounter{page}{0}
\begin{flushright}
hep-th/0702106\\
%ITP--UH--/\\
\end{flushright}

\vskip 2.0cm

\begin{center}

{\Large\bf Sigma Models with $\Ncal{=}8$ Supersymmetries \\[2mm]
in 2+1 and 1+1 Dimensions
}

\vspace{15mm}
{\Large Alexander D. Popov}
\\[5mm]
\noindent {\em Institut f\"ur Theoretische Physik,
Leibniz Universit\"at Hannover \\
Appelstra\ss{}e 2, 30167 Hannover, Germany }
\\[2mm]
and
\\[2mm]
\noindent {\em Bogoliubov Laboratory of Theoretical Physics, JINR\\
141980 Dubna, Moscow Region, Russia}
\\[5mm]
{Email: {\tt popov@itp.uni-hannover.de}}

\vspace{15mm}

\begin{abstract}
\noindent
We introduce an $\Ncal{=}8$ supersymmetric extension of the Bogomolny-type model 
for Yang-Mills-Higgs fields in 2+1 dimensions related with twistor string theory.
It is shown that this model is equivalent to an $\Ncal{=}8$ supersymmetric U($n$)
chiral model in 2+1 dimensions with a Wess-Zumino-Witten-type term. Further 
reduction to 1+1 dimensions yields $\Ncal{=}(8,8)$ supersymmetric extensions 
of the standard U($n$) chiral model and Grassmannian sigma models.
\end{abstract}

\end{center}
\end{titlepage}

\section{Introduction and Summary}

\noindent
Nonlinear sigma models in $k$ dimensions describe mappings of a $k$-dimensional
manifold $X$ into a manifold $Y$ (target space). In particular, as target spaces 
one can consider Lie groups $G$ (chiral models) and homogeneous spaces $G/H$ for 
closed subgroups $H\subset G$. Sigma models and their 
$\Ncal$-extended supersymmetric generalizations play an important role both in 
physics and mathematics (see e.g.~\cite{Zakr, Perelomov}). For instance, 
two-dimensional sigma models serve as a theoretical laboratory for the study of 
more complicated (quantum) super Yang-Mills theory since they share many of 
its features such as asymptotic freedom, nontrivial topological structure, 
the existence of instantons, ultraviolet finiteness for the $\Ncal{=}4$ 
supersymmetric case etc.~\cite{Pol}. Moreover, supersymmetric two-dimensional 
sigma models are the building blocks for superstring theories~\cite{Pol, GSW}.

Recall that for two-dimensional nonlinear sigma models admitting a Lagrangian 
formulation the number of supersymmetries is intimately related to the geometry
of the target space. Namely, it was argued that Lagrangian $\Ncal{=}1$ models 
can be defined for any target space $Y$, for $\Ncal{=}2$ the target space must 
be K\"ahler, for $\Ncal{=}4$ it must be hyper-K\"ahler, and no Lagrangian models 
were introduced for $\Ncal{>}4$~\cite{Zumino, AGF}. Similar results hold for 
sigma models in three dimensions. In particular, this means that a target space 
$Y$ admits no more than $\Ncal{=}1$ supersymmetry in the case of (non-K\"ahler) 
group manifolds $G$ and $\Ncal{\le}2$ supersymmetries for homogeneous K\"ahler 
spaces $G/H$.

The field equations of the standard $G$ and $G/H$ sigma models in 1+1 and 2+0
dimensions can be obtained by dimensional reduction of 
the self-dual Yang-Mills (SDYM) equations in 2+2 dimensions, with a gauge group 
$G$~\cite{Ward}. Concretely, the SDYM model reduced to two dimensions is 
{\it equivalent} to the sigma model with $G$-valued scalar fields, while the 
$G/H$ sigma model arises after imposing additional algebraic constraints.
Similar reduction to 2+1 dimensions yields a modified integrable chiral 
model~\cite{Ward88}. Recall that the SDYM model in 2+2 dimensions can be endowed
with up to four supersymmetries~\cite{Sieg, GNK}. Reducing the $\Ncal$-extended
supersymmetric SDYM equations in 2+2 dimensions to 2+1 and 1+1 dimensions yields
models which have twice as many supersymmetries (cf.~\cite{Seib} for reductions 
from 3+1 dimensions). We will show that for $G{=}$U($n$) and $\Ncal{=}4$ these 
models are equivalent to U($n$) chiral models with $\Ncal{=}8$ supersymmetries. 
These new supersymmetric sigma models in 2+1 and 1+1 dimensions are well 
defined on the level of equations of motion, but their Lagrangian formulation 
is not known yet.

In this note we concentrate on the reduction of the $\Ncal{=}4$ SDYM equations 
(instead of arbitrary $\Ncal{\le}4$) in 2+2 dimensions since for this case a 
Lagrangian can be written down at least in terms of the component fields of a 
reduced Yang-Mills-type supermultiplet. Moreover, it was shown by 
Witten~\cite{Wit} that the $\Ncal{=}4$ SDYM model appears in twistor string 
theory, which is a B-type topological string with the supertwistor space 
$\C P^{3|4}$ as a target space\footnote{For other variants of twistor string 
models see~\cite{group3}.}. This fact gives additional arguments in favour of
introducing $\Ncal=8$ supersymmetric sigma models in 2+1 and 1+1 dimensions 
related with twistor string theory and of studying their properties.

\vspace{5mm}

\section{$\Ncal{=}4$ supersymmetric SDYM equations in 2+2 dimensions}

\noindent
{\bf Superspace $\R^{4|16}$.} Let us consider the four-dimensional space
$\R^{2,2}:=(\R^4, g)$ with the metric
\begin{equation}\label{s2}
\diff s^2 = g_{\mu\nu}\diff x^\mu\diff x^\nu =\det (\diff x^{\a\da})=
\diff x^{1\1}\diff x^{2\2} -
\diff x^{2\1}\diff x^{1\2}
\end{equation}
with $(g_{\mu\nu})=$diag$(-1,+1,+1,-1)$. Here $\mu , \nu , ...=1,...,4$ are
vector indices and $\a = 1,2$, $\da = \1 , \2 $ are spinor indices. We
choose the real coordinates\footnote{Our conventions are chosen to match those
of~\cite{LPS2} after reduction to the space $\R^{2,1}$ with coordinates
$(t,x,y)$.} $(x^\mu ) = (x^a , \tilde t) =(t,x,y,\tilde t)$ with $a,b,...=1,2,3$
such that
\begin{equation}\label{iso}
x^{1\1} = \sfrac12(t-y),\ x^{1\2} = \sfrac12(x+\tilde t),\ x^{2\1} = \sfrac12(x-
\tilde t)\quad\mbox{and}\quad x^{2\2} = \sfrac12(t + y)  \ .
\end{equation}

On the space $\R^{2,2}$ one can introduce real Majorana-Weyl spinors and 
extend $\R^{2,2}$ to a space with additional anticommuting (Grassmann) 
coordinates $\th^{i\a}$ and $\h_i^{\da}$ of helicity $+\sfrac12$ and 
$-\sfrac12$, respectively. Here index $i=1,...,4$ parametrizes fundamental
and its conjugate representations of the R-symmetry group 
SL(4, $\R$)~\cite{Sieg}. Thus, $(x^{\a\da}, \h_i^{\da}, \th^{i\a})$ are 
coordinates on superspace $\R^{4|16}$.

\noindent
{\bf Supersymmetry algebra.} The $\Ncal{=}4$ supersymmetry algebra in 2+2 
dimensions is generated by
$P_{\a\da}=\pa_{\a\da}=\pa / \pa x^{\a\da}$ and 16 real supercharges
\begin{equation}\label{2.12}
Q_{i\a}:=\pa_{i\a}-\h_i^{\da}\pa_{\a\da}\quad\mbox{and}
\quad Q^{i}_{\da}:=\pa^i_{\da}-\th^{i\a}\pa_{\a\da}\ ,
\end{equation}
with $\pa_{i\a}:={\pa}/{\pa \th^{i\a}}$ and 
$\pa^i_{\da}:={\pa}/{\pa \h^{\da}_i}$.
The only nontrivial (anti)commutators in this superalgebra read
\begin{equation}
\{Q_{i\a}, Q^{j}_{\da}\} = -2\de_i^j \pa_{\a\da}.
\end{equation}

In what follows we will also need superderivatives
\begin{equation}
D_{i\a}:=\pa_{i\a}+\h_i^{\da}\pa_{\a\da}\quad\mbox{and}
\quad D^{i}_{\da}:=\pa^i_{\da}+\th^{i\a}\pa_{\a\da}\ ,
\end{equation}
which anticommute with the operators (\ref{2.12}) and satisfy
\begin{equation}
\{D_{i\a}, D^{j}_{\db}\} = 2\de_i^j \pa_{\a\db}\ .
\end{equation}

\smallskip

\noindent
{\bf Antichiral superspace.} On the superspace $\R^{4|16}$ we can introduce
spin-tensor fields depending on both bosonic and fermionic coordinates 
(superfields) and impose on them various constraints. In particular, on any 
superfield $\Acal$ one can impose the so-called antichirality conditions 
$\Lcal_{D_{i\a}}\Acal =0$, where $\Lcal_Z$ denotes the Lie derivative along a 
vector superfield $Z$. One can easily solve these equations by using a 
coordinate transformation on superspace $\R^{4|16}$,
\begin{equation}\label{xht}
(x^{\a\da},\ \h^{\da}_i,\ \th^{i\a})\ \to\ (\tilde x^{\a\da}=x^{\a\da}{-}
\th^{i\a}\h_i^{\da},\ \h^{\da}_i,\ \th^{i\a})\ ,
\end{equation}
under which $\pa_{\a\da}, D_{i\a}$ and $D^i_{\da}$ transform to the operators
\begin{equation}\label{pDD}
\tilde\pa_{\a\da}=\pa_{\a\da}\ , \quad \tilde D_{i\a}=\pa_{i\a}\ , \quad 
\tilde D^i_{\da}=\pa^i_{\da}+ 2\th^{i\a}\pa_{\a\da}\ .
\end{equation}
The antichirality conditions then mean that a superfield $\Acal$ satisfies the 
equations
\begin{equation}\label{xh}
\tilde D_{i\a}\Acal =0
\end{equation}
meaning that $\Acal$ is defined on superspace $\R^{4|8}\subset\R^{4|16}$
called antichiral superspace with coordinates $(x^{\a\da},\ \h_i^{\da})$.
Note that for transformed supercharges we have
\begin{equation}\label{2.10}
\tilde Q_{i\a}=\pa_{i\a}-2\h_i^{\da}\pa_{\a\da}\quad\mbox{and}
\quad \tilde Q^{i}_{\da}=\pa^i_{\da}\ .
\end{equation}
In the following we will often omit the tilde when dealing with the antichiral 
superspace.

\smallskip

\noindent
{\bf $\Ncal{=}4$ SDYM in superfields.}
The field content of $\Ncal{=}4$ supersymmetric SDYM is given by a supermultiplet
$(A_{\a\da}, \chi^{i\a}, \phi^{ij}, \tilde\chi^{\da}_i, G_{\da\db})$ of fields
on $\R^{2,2}$ of helicities $(+1, +\sfrac12, 0, -\sfrac12, -1)$. Here $A_{\a\da}$
are the components of a gauge potential with the field strength
$F_{\a\da ,\b\db}{=}\pa _{\a\da}A_{\b\db}{-}\pa_{\b\db} A_{\a\da}{+}
[A_{\a\da}, A_{\b\db}]$.
Note that the scalars $\phi^{ij}$ are antisymmetric in $ij$ and all the fields,
including the fermionic ones  $\chi^{i\a}$ and $\tilde\chi^{\da}_i$, live in the
adjoint representation of the gauge group U($n$).

The $\Ncal =4$ SDYM equations~\cite{SemVol, Sieg} can be written in terms 
of superfields on antichiral superspace $\R^{4|8}$~\cite{Sieg, DevOg}. 
Namely, all fields from the above $\Ncal =4$ supermultiplet can be combined 
into superfields $\Acal_{\a\da}$ and $\Acal^i_{\da}$
on $\R^{4|8}$ in terms of which the  $\Ncal =4$ SDYM equations read
\begin{equation}\label{sdym1}
[\nabla_{\a\da},\nabla_{\b\db}]+[\nabla_{\a\db},\nabla_{\b\da}]=0\ ,\quad
[\nabla_{\da}^i,\nabla_{\b\db}]+[\nabla_{\db}^i,\nabla_{\b\da}]=0\ ,\quad
\{\nabla_{\da}^i,\nabla_{\db}^j\}+\{\nabla_{\db}^i, \nabla_{\da}^j\}=0\ ,
\end{equation}
where we have introduced the covariant derivatives
\begin{equation}\label{covder}
\nabla_{\alpha\da}:=\pa_{\alpha\da} +\Acal_{\alpha\da}\quad\mbox{and}\quad
\nabla_{\da}^i:=\pa_{\da}^i+\Acal_{\da}^i\ .
\end{equation}
Note that (\ref{sdym1}) can be combined into the manifestly supersymmetric equations
\begin{equation}\label{sdym3}
\{\tilde\nabla^i_{\da}, \tilde\nabla^j_{\db}\} +
\{\tilde\nabla^i_{\db}, \tilde\nabla^j_{\da}\}=0
\end{equation}
with
\begin{equation}
\tilde\nabla^i_{\da}:=\nabla^i_{\da} + 2\th^{i\a}\nabla_{\alpha\da}=
\tilde D^i_{\da} + \tilde\Acal^i_{\da}\quad\mbox{and}\quad
\tilde\Acal^i_{\da}:= \Acal^i_{\da} + 2\th^{i\a}\Acal_{\a\da}\ ,
\end{equation}
where $\Acal_{\a\da}$ and $\Acal^i_{\da}$ depend only on $x^{\a\da}$
and $\h_i^{\da}$.

It is not difficult to see that equations (\ref{sdym3}) are the compatibility
conditions for the linear system of differential equations
\begin{equation}\label{ls}
\la_\pm^{\da}(\tilde D^i_{\da} + \tilde\Acal^i_{\da})\psi_\pm =0\ ,
\end{equation}
where $\la_\pm^{\da}=\ve^{\da\db}\la^\pm_{\db},\ (\la^+_{\db})=(1\ \la_+)^\top ,
\ (\la^-_{\db})=(\la_-\ 1)^\top$ and the extra (local) coordinates $\la_\pm$
lie on patches $U_\pm$ covering the Riemann sphere $\C P^1= U_+\cup U_-$ (see
e.g.~\cite{PoSa}). Here $\psi_\pm$ are $n\times n$ matrices depending not
only on $x^{\a\da}$ and $\eta_i^{\da}$ but also (holomorphically) on 
$\la_\pm\in U_\pm$.

The field equations of the $\Ncal =4$ SDYM model in the component fields read
\begin{subequations}\label{sdym2}
\begin{eqnarray}
F_{{\da}{\db}}=0\ ,\quad D_{{\a}{\da}}\,\chi^{i\a}=0\ ,\quad
D_{{\a}{\da}}\,D^{{\a}{\da}}\phi^{ij} + \{\chi^{i\a},\,\chi^j_{\a}\}=0\ ,
\\
D_{{\a}{\da}} \tilde{\chi}^{\da}_i + [\chi_{\a}^{j},\ \phi_{ij} ]=0\ ,\quad
\ve^{{\da}\dot\g}D_{\a{\da}}G_{\dot\g\db}-\sfrac12 \{\chi_{\a}^{i},\,
\tilde{\chi}_{i\db}\}-\sfrac{1}{4}[\phi_{ij},\, D_{\a{\db}}\phi^{ij}]=0\ ,
\end{eqnarray}
\end{subequations}
where $F_{{\da}{\db}}:=-\sfrac12\ve^{\a\b}F_{\a{\da}, \b{\db}}$, 
$D_{{\a}{\da}} :=\pa_{{\a}{\da}} +[A_{{\a}{\da}},\, \cdot\, ]$ and 
$\phi_{ij}:=\sfrac{1}{2!}\ve_{ijkl}\phi^{kl}$.
These equations can be extracted from (\ref{sdym1}) by using $\h$-expansions and Bianchi identities (see e.g.~\cite{DevOg}). We will not reproduce this 
derivation. Note only that (\ref{sdym2}) follows from the 
Lagrangian~\cite{Sieg, Wit}
\begin{equation}\label{2.18}
{\cal L}=\tr \left (G^{{\da}{\db}} F_{{\da}{\db}}+ 
\tilde\chi_i^{\da}D_{{\a}{\da}}\chi^{i\a}+
\phi_{ij}D_{{\a}{\da}}D^{{\a}{\da}}\phi^{ij}+ \phi_{ij}\chi^{i\a}\chi^j_{\a}
\right )\ .
\end{equation}

\vspace{5mm}

\section{$\Ncal{=}8$ supersymmetric sigma models in 2+1 dimensions}

\noindent
{\bf Reduction and spinors on $\R^{2,1}$.} The $\Ncal{=}8$ supersymmetric 
Bogomolny-type equations in 2+1 dimensions are obtained from the described 
$\Ncal{=}4$ super SDYM equations by the dimensional reduction 
$\R^{2,2}\to\R^{2,1}$. Namely, we impose the $\pa_4$-invariance condition on 
all the fields $(A_{\a\da},\ \chi^{i\a},\ \phi^{ij},\ \tilde\chi^{\da}_i ,\ 
G_{\da\db})$ from the $\Ncal{=}4$ supermultiplet. Also, the components $A_\m$ 
of a gauge potential split into the components $A_a$ in 2+1 dimensions and the 
Lie-algebra valued scalar field $\vp :=A_4$ (Higgs field). To see how
this splitting looks in spinor notation, we briefly discuss spinors in 2+1 
dimensions.

Recall that $\Ncal{=}4$ SDYM theory on $\R^{2,2}$ has SL(4, $\R)\cong$ Spin(3,3) 
as an R-symmetry group~\cite{Sieg}. Analogously to the case of standard 
$\Ncal{=}4$ super Yang-Mills (SYM) in Minkowski space with the Spin(6) 
R-symmetry, the appearance of the group Spin(3,3) can be interpreted
via a reduction of $\Ncal{=}1$ SYM theory on space $\R^{5,5}\cong\R^{2,2}
\times\R^{3,3}$ 
to $\R^{2,2}$ with internal space $\R^{3,3}$~\cite{GNK}. Furthermore, after 
reduction from $\R^{2,2}$ to $\R^{2,1}$ the R-symmetry group becomes Spin(4,4) 
and supersymmetry gets enlarged to $\Ncal=8$ with Spin(4,3) as the manifest 
R-symmetry group (cf.~\cite{Seib} for Minkowski and~\cite{BT} for Euclidean 
signatures). Roughly speaking, this happens due to no distinction between dotted 
and undotted spinor indices in three dimensions. Recall that the rotation group 
SO(2,2) of $\R^{2,2}$ is locally isomorphic to
SU(1,1)$_L\times$SU(1,1)$_R\cong$Spin(2,1)$_L\times$Spin(2,1)$_R\cong$Spin(2,2). 
Upon dimensional reduction to 2+1 dimensions, the rotation group of 
$\R^{2,1}=(\R^3, g)$ with $g=(g_{ab})=$diag($-$1,+1,+1)
is locally SU(1,1)$\cong$Spin(2,1), which is the diagonal subgroup of
Spin(2,1)$_L\times$ Spin(2,1)$_R\cong$ Spin(2,2).
Therefore, the distinction between dotted and undotted indices disappear.

\smallskip

\noindent
{\bf Coordinates and derivatives on $\R^{3|16}$.} The $\pa_4$-invariance reduces
superspace  $\R^{4|16}$ with coordinates $x^\mu , \h_i^{\da}$ and $\th^{i\a}$ to
$\R^{3|16}$ with coordinates $x^a$, $\h_i^{\a}$ and $\th^{i\a}$. Furthermore,
$x^a$ and $\h_i^{\a}$ parametrize reduced antichiral superspace  $\R^{3|8}$. For
bosonic coordinates $x^{\a\db}\to x^{\a\b}$ in spinor notation we have
\begin{equation}\label{iso2}
x^{\a\b}=\sfrac12(x^{\a\b}+x^{\b\a})+\sfrac12(x^{\a\b}-x^{\b\a})=
x^{(\a\b)}+x^{[\a\b]}\ .
\end{equation}
Thus, we have coordinates
\begin{equation}\label{iso3}
y^{\a\b}:=x^{(\a\b)}\quad\mbox{with}\quad y^{11}=x^{11}=\sfrac12(t-y),\
y^{12}=\sfrac12(x^{12}+x^{21})=\sfrac12 x,\  y^{22}=x^{22}=\sfrac12(t+y)
\end{equation}
on $R^{2,1}$ and  $x^{[\a\b]}= -\ve^{\a\b}x^4 = -\ve^{\a\b}\tilde t$,
where $\ve^{12}= -\ve^{21}=1$.

For derivatives we obtain
\begin{equation}\label{pdadb}
\pa_{\a\b}= \sfrac12 (\pa_{\a\b}+\pa_{\b\a})+
\sfrac12 (\pa_{\a\b}-\pa_{\b\a})=\pa_{(\a\b)}-
\ve_{\a\b}\pa_4=\pa_{(\a\b)}-\ve_{\a\b}\pa_{\tilde t}\ ,
\end{equation}
where $\ve_{12}=-\ve_{21}=-1$ and
\begin{equation}\label{p1212}
\pa_{(11)}=\frac{\pa}{\pa y^{11}}=\pa_{t}-\pa_y\ ,\quad
\pa_{(12)}=\pa_{(21)}=\frac{1}{2}\frac{\pa}{\pa y^{12}}=\pa_x\ ,\quad
\pa_{(22)}=\frac{\pa}{\pa y^{22}}=\pa_{t}+\pa_y\  .
\end{equation}
For the operators (\ref{pDD}) acting on $\tilde t$-independent superfields
we have
\begin{equation}\label{hatDs}
\hat D_{i\a} =\pa_{i\a}\quad\mbox{and}\quad
\hat D_{\a}^i =\pa_{\a}^i+2\th^{i\b}\pa_{(\a\b )}\ .
\end{equation}
Similarly, supercharges (\ref{2.10}) reduce to the operators
\begin{equation}\label{hatQs}
\hat Q_{i\a} =\pa_{i\a}-2\h_i^{\b}\pa_{(\a\b )}\quad\mbox{and}\quad
\hat Q_{\a}^i =\pa_{\a}^i\ ,
\end{equation}
anticommuting with (\ref{hatDs}).

\noindent
{\bf $\Ncal=8$ supersymmetric Bogomolny-type equations on $\R^{2,1}$.}
 After imposing the condition of $\tilde t$-independence on all
fields in the linear system (\ref{ls}), we obtain the equations
\begin{equation}\label{ls5}
\la^{\a}_\pm (\hat D_{\a}^i + \hat\Acal_{\a}^i)\psi_\pm =0
\end{equation}
with
\begin{equation}\label{3.8}
\hat\Acal^i_{\a}= \Acal^i_{\a} + 2\th^{i\b}(\Acal_{(\a\b)}-
\ve_{\a\b}\tilde\vp )\ ,
\end{equation}
and  $\hat D_{\a}^i$ given in (\ref{hatDs}). Here $\Acal^i_{\a}$, 
$\Acal_{(\a\b)}$ and $\tilde\vp$ are superfields depending only on 
$y^{\a\b}$ and $\h_i^{\b}$.

The compatibility conditions for the linear system (\ref{ls5}) read
\begin{equation}\label{3.9}
\{\hat D^i_{\a}+\hat\Acal_{\a}^i\ ,\ \hat D^j_{\b}+\hat\Acal_{\b}^j\} +
\{\hat D^i_{\b}+\hat\Acal_{\b}^i\ ,\ \hat D^j_{\a}+\hat\Acal_{\a}^j\} = 0\ .
\end{equation}
As usual, these manifestly $\Ncal=8$ supersymmetric equations are equivalent 
to equations in component fields,
\begin{subequations}\label{3.10}
\begin{eqnarray}
&&f_{{\a}{\b}}+D_{{\a}{\b}}\vp =0\ , \quad
D_{{\a}{\b}}\,\chi^{i\b} + \ve_{{\a}{\b}}\,[\vp ,\, \chi^{i\b} ]=0\ ,
\\
&&D_{{\a}{\b}}\,D^{{\a}{\b}}\phi^{ij} + 2[\vp ,\, [\vp,\phi^{ij}]]+
\{\chi^{i\a},\,\chi^j_{\a}\}=0\ ,
\\
&&D_{{\a}{\b}}\ \tilde{\chi}^{\b}_i-\ve_{{\a}\b}\,[\vp,\,
\tilde{\chi}^{\b}_i]+[\chi_{\a}^{j},\ \phi_{ij} ]=0\ ,
\\
&&\ve^{{\g}{\de}}D_{{\a}{\g}}G_{{\de}{\b}}
{+}[\vp,\, G_{\a\b}]- {\sfrac12}\{\chi_{\a}^i,\tilde{\chi}_{i\b}\}-
\sfrac{1}{4}[\phi_{ij},D_{{\a}{\b}}\phi^{ij}]-
\sfrac14\ve_{{\a}{\b}}[\phi_{ij},[\phi^{{ij}},\vp ]]=0 ,
\end{eqnarray}
\end{subequations}
where $D_{{\a}{\b}}:=\pa_{({\a}{\b})}+[A_{({\a}{\b})},\ \cdot\ ],\   
f_{{\a}{\b}}:=-\sfrac12\ve^{{\g}{\de}}[D_{{\a}{\g}}, 
D_{{\b}{\de}}]$ and $\vp:=A_4=A_{\tilde t}$. Obviously, these equations 
are $\pa_4$-reduction of (\ref{sdym2}).

\smallskip

\noindent
{\bf Supersymmetric sigma models.} Note that matrices $\psi_\pm$ in (\ref{ls5})
are defined up to a gauge transformation generated by a matrix which does not
depend on $\la_\pm$ and therefore one can choose a gauge such that
\begin{equation}\label{3.11}
\j_+=\Phi^{-1}+{\cal O}(\la_+ )\quad\mbox{and}\quad
\j_-={\mathbf 1}_n +\la_-\Ups +{\cal O}(\la_-^{2} )\ ,
\end{equation}
where $\Phi$ is a U($n$)-valued superfield and $\Ups$ is a $u(n)$-valued 
superfield both depending only on $y^{{\a}{\b}}$ and $\h_i^{\a}$. For this 
gauge, from (\ref{ls5}) we obtain
\begin{equation}\label{3.12}
\hat\Acal^i_{1}=0 \quad\mbox{and}\quad\hat\Acal^i_{2}=
\Phi^{-1}\hat D^i_{2}\Phi \ ,
\end{equation}
and from (\ref{3.8}) we have
\begin{subequations}\label{3.13}
\begin{eqnarray}
&\Acal^i_{1}=0 \ ,\quad \Acal_{(11)}=0\ ,\quad \Acal_{(12)}-\tilde\vp =0\ ,\\
&\Acal^i_{2}=\Phi^{-1}\pa^i_{2}\Phi\ ,\quad
\Acal_{(12)}+\tilde\vp =\Phi^{-1}\pa_{(12)}\Phi  \quad\mbox{and}\quad
\Acal_{(22)}=\Phi^{-1}\pa_{(22)}\Phi \ ,
\end{eqnarray}
\end{subequations}

Substituting  (\ref{3.12}) into (\ref{3.9}), we obtain equations
\begin{equation}\label{3.14}
\hat D_{1}^i(\Phi^{-1}\hat D_{2}^j\ \Phi )+ \hat D_{1}^j(\Phi^{-1}\hat D_{2}^i\ 
\Phi )=0
\end{equation}
which after using (\ref{hatDs}) and (\ref{3.13}) read
\begin{equation}\label{3.15}
\pa_x(\Phi^{-1}\pa_x\Phi )+\pa_y(\Phi^{-1}\pa_y\Phi )-\pa_t(\Phi^{-1}\pa_t\Phi )+
\pa_y(\Phi^{-1}\pa_t\Phi )-\pa_t(\Phi^{-1}\pa_y\Phi )=0\ ,
\end{equation}
\begin{equation}\label{3.16}
\pa_{1}^i(\Phi^{-1}\pa_x\Phi ){-}\pa_t(\Phi^{-1}\pa_{2}^i\Phi ){+}
\pa_y(\Phi^{-1}\pa_{2}^i\Phi ){=}0 ,\
\pa_{1}^i(\Phi^{-1}\pa_t\Phi ){+}\pa_{1}^i(\Phi^{-1}\pa_{y}\Phi ){-}
\pa_x(\Phi^{-1}\pa_{2}^i\Phi ){=}0\ ,
\end{equation}
\begin{equation}\label{3.17}
\pa_{1}^i(\Phi^{-1}\pa_{2}^j\Phi )+\pa_{1}^j(\Phi^{-1}\pa_{2}^i\Phi )=0\ .
\end{equation}
Note that the last two terms in (\ref{3.15}) are the Wess-Zumino-Witten terms 
which spoil the standard Lorentz invariance but yield  an integrable U($n$) 
chiral model in 2+1 dimensions. For reduction to 1+1 dimensions one should 
simply put $\pa_y\Phi =0$ in (\ref{3.15})-(\ref{3.17}) obtaining an $\Ncal{=}8$
supersymmetric extensions of the standard U($n$) chiral model in two dimensions
with field equations
\begin{subequations}\label{3.18}
\begin{eqnarray}
&\pa_t(\Phi^{-1}\pa_t\Phi )-\pa_x(\Phi^{-1}\pa_x\Phi )=0\ ,\quad
\pa_{1}^i(\Phi^{-1}\pa_{2}^j\Phi )+\pa_{1}^j(\Phi^{-1}\pa_{2}^i\Phi )=0\ ,\\
&\pa_{1}^i(\Phi^{-1}\pa_x\Phi ){-}\pa_t(\Phi^{-1}\pa_{2}^i\Phi )=0\ ,\quad
\pa_{1}^i(\Phi^{-1}\pa_t\Phi )-\pa_x(\Phi^{-1}\pa_{2}^i\Phi )=0\ .
\end{eqnarray}
\end{subequations}
For $\Phi$ taking values in the Grassmannian manifold Gr$(k,n){\subset}\,$U($n$), 
equations (\ref{3.15})-(\ref{3.17}) and (\ref{3.18}) describe correspondingly supersymmetric Grassmannian sigma models in 2+1 and 1+1 dimensions.

There is not yet a Lagrangian description of equations (\ref{3.15})-(\ref{3.17}) 
or (\ref{3.18}). However, using the equivalence of equations (\ref{3.10}) to 
(\ref{3.14}), one can write explicitly a Lagrangian in terms of fields 
$(A_{({\a\b})},\ \chi^{i\a},\ \vp,\ \phi^{ij},\ \tilde\chi^{\a}_i ,\
G_{\a\b})$. The proper Lagrangians follow from (\ref{2.18}) by reduction to 
2+1 and 1+1 dimensions. It is a challenging task to find Lagrangians in 
terms of the U($n$)-valued superfield $\Phi$.

\smallskip

\noindent
{\bf Supersymmetry transformations.} For brevity, we consider only 2+1 
dimensions, where the 16 supercharges have the form (\ref{hatQs}). Further 
reduction to 1+1 dimensions does not create any problem. From
(\ref{hatQs}) we obtain
\begin{equation}\label{3.19}
\{\hat Q_{i\a}, \hat Q^j_{\b}\}=-2\de^j_i\pa_{(\a\b)}\ .
\end{equation}
On a (scalar) superfield $\Sigma$ an infinitesimal supersymmetry transformation 
$\hat\de$ acts by 
\begin{equation}\label{3.20}
\hat\de\Sigma :=\eps^{i\a}\hat Q_{i\a}\Sigma + \eps_i^{\a}\hat Q^i_{\a}\Sigma\ ,
\end{equation}
where $\eps^{i\a}$ and $\eps_i^{\a}$ are 16 Grassmann parameters. In particular, 
for coordinates $y^{\a\b}$ and $\h^{\b}_i$ on the antichiral superspace 
$\R^{3|8}$ we have $\hat\de y^{\a\b}=-2\eps^{i(\a}\h^{\b )}_i$ and 
$\hat\de\h^{\a}_i=\eps^{\a}_i$.

It is obvious that the sigma model field equations (\ref{3.14}) are invariant 
under the supersymmetry transformations (\ref{3.20}) because the operators 
$\hat D^i_{\a}$ as well as $\hat D_{i\a}$ anticommute with the supersymmetry
generators $\hat Q_{i\a}$ and $\hat Q^j_{\b}$. Note that these $\Ncal=8$ 
supersymmetric extensions of the U($n$) and 
Gr($k,n$)=U($n$)/U($k$)$\times$U($n{-}k$) sigma models in 2+1 and 1+1 
dimensions are not the standard ones defined only for $\Ncal\le 1$ and 
$\Ncal\le 2$, respectively. It will be interesting to study this new kind of 
sigma models in more detail.

\bigskip

\section*{Acknowledgements}

\noindent
The author would like to thank Olaf Lechtenfeld for reading the manuscript
and useful remarks. This work was supported in part by the Deutsche
 Forschungsgemeinschaft~(DFG).

\end{document}